\documentclass[aps,prl,preprint,superscriptaddress]{revtex4}
\usepackage{amsfonts}
\usepackage{amsmath}
\usepackage{graphicx}
\usepackage{dcolumn}
\usepackage{bm}

\input{tcilatex}
\begin{document}

\title{Metriplectic Framework for\\
Dissipative Magneto-Hydrodynamics}
\author{Massimo Materassi}
\email{massimo.materassi@fi.isc.cnr.it}
\affiliation{ISC-CNR (Istituto dei Sistemi Complessi, Consiglio Nazionale delle
Ricerche),\ via Madonna del Piano 10, I-50019 Sesto Fiorentino, Italy}
\author{Emanuele Tassi}
\email{tassi.emanuele@cpt.univ-mrs.fr}
\affiliation{Centre de Physique Th\'eorique, CNRS -- Aix-Marseille Universit\'es, Campus
de Luminy, case 907, F-13288 Marseille cedex 09, France}
\date{\today }

\begin{abstract}
The metriplectic framework, which permits to formulate an algebraic
structure for dissipative systems, is applied to visco-resistive
Magneto-Hydrodynamics (MHD), adapting what had already been done for
non-ideal Hydrodynamics (HD). The result is obtained by extending the HD
symmetric bracket and free energy to include magnetic field dynamics and
resistive dissipation. The correct equations of motion are obtained once one
of the Casimirs of the Poisson bracket for ideal MHD is identified with the
total thermodynamical entropy of the plasma. The metriplectic framework of
MHD is shown to be invariant under the Galileo Group. The metriplectic
structure also permits to obtain the asymptotic equilibria toward which the
dynamics of the system evolves. This scheme is finally adapted to the
two-dimensional incompressible resistive MHD, that is of major use in many
applications.

\textbf{CORRESPONDING AUTHOR:}

Massimo Materassi, ISC-CNR (Istituto dei Sistemi Complessi, Consiglio
Nazionale delle Ricerche),\ via Madonna del Piano 10, I-50019 Sesto
Fiorentino, Italy, tel. +39-347-6113002, fax: +39-055-5226683, e-mail:
massimo.materassi@fi.isc.cnr.it
\end{abstract}

\pacs{(...)}
\maketitle

\section{Introduction}

The impossibility of solving analytically the overwhelming majority of
differential equations in Physics soon convinced physicists to investigate
the properties of dynamical systems without searching for all the possible
solutions. Integral properties of the solutions were then investigated, as
conserved quantities, and not much more than the esthetic taste inspired
theorists to formulate those shortcuts in a mathematically cleaner way: this
is more or less the history of Action Principles \cite{goldstein,basdevant},
beginning as acute observations on special problems, and soon generating the
wonderful offspring of Lagrangian Dynamics (with its noble descendants of
path integral representations \cite{Feynman.Book}), and Hamiltonian Dynamics.

Algebrization of dynamical systems appears to be the final destination of
that virtuous route \cite{Arnold.Book}: in the Hamiltonian framework
dynamics is turned into a bracket algebra of observable quantities, and then
physical properties of systems, especially in terms of conserved quantities
and symmetries \cite{Symplectization}, can be obtained without even the need
of going back to the equations of motion \cite{Landau.Mechanics}.
Hamiltonian dynamics has, also, represented a huge breakthrough to Quantum
Physics \cite{Dirac.Book}, that is exquisitely an algebraic formulation.

This cultural and methodological evolution, starting with some symmetry
observations and ending up with the bracket algebr\ae , appears to be
natural for \textit{conservative systems}.

A very promising strategy to algebrize the dynamics of a dissipative system
is the \textit{metriplectic framework} \cite{Morrison84.001,Mor09}. The
system at hand must be \textit{complete}, i.e. one must be able to keep
trace of the total energy during the motion:\ typically, this means
including all the energy exchanges in a conserved Hamiltonian. In other
words, the metriplectic framework is applicable to closed systems.

Dissipation is generally understood as the interaction of dynamical
variables of the otherwise Hamiltonian system with other \textit{%
microscopic, statistically treated, degrees of freedom} (MSTDOF), giving
rise to friction. The system is extended to include the MSTDOF, and this
closes the system. The dynamics of the closed system with friction is then
assigned by defining a symmetric extension of the Poisson bracket algebra,
and an extension of the Hamiltonian to free energy. In order to extend the
Hamiltonian to the free energy of the closed system, the entropy $S$ of the
MSTDOF will be used.

Hamiltonian dynamics evolves any quantity $f$ as $\dot{f}=\left\{
f,H\right\} $, being $\left\{ f,g\right\} $ the Poisson bracket, while
no-friction condition would imply no entropy production in the Hamiltonian
system. Then, the entropy $S$ must be conserved in the Hamiltonian limit of
the dissipative system: $\left\{ S,H\right\} =0$. For noncanonical
Hamiltonian systems, $S$ is then expected to be expressable through \textit{%
Casimir functionals of the Poisson bracket} $\left\{ f,g\right\} $, i.e.
quantities $C$ such that%
\begin{equation}
\begin{array}{ccc}
\left\{ C,f\right\} =0 & \forall & f.%
\end{array}
\label{nonid.MHD.0009.Casimir}
\end{equation}%
The Hamiltonian is hence extended by defining the \textit{free energy}%
\begin{equation}
F=H+\lambda C.  \label{nonid.MHD.0011.mp2}
\end{equation}%
The coefficient $\lambda $ in (\ref{nonid.MHD.0011.mp2}) is a constant:
under the hypothesis of thermal equilibrium for MSTDOF and asymptotic
equilibrium for the system, this $\lambda $ will coincide with minus the
temperature of MSTDOF, but in general it should be understood just as an
arbitrary constant coefficient left indicated.

The framework is completed by prescribing that the evolution of any quantity 
$f$ is generated by $F$ via an extension $\left\langle \left\langle
f,g\right\rangle \right\rangle $ of the original Poisson bracket $%
\left\langle \left\langle f,g\right\rangle \right\rangle =\left\{
f,g\right\} +\left( f,g\right) $, where the symbol $\left( f,g\right) $ is
symmetric, bilinear and semi-definite \cite{morrison.paradigm}. For
instance, for the positive semi-definite case, we have:%
\begin{equation*}
\begin{array}{cccc}
\left( f,g\right) =\left( g,f\right) , & \left( f,f\right) \geq 0 & \forall
& f,g.%
\end{array}%
\end{equation*}%
In a metriplectic framework the evolution is then generated as:%
\begin{equation}
\dot{f}=\left\langle \left\langle f,F\right\rangle \right\rangle
\label{nonid.MHD.0014.mp1}
\end{equation}%
(the symmetric bracket $\left( f,g\right) $ will be defined so to cancel out
the presence of the coefficient $\lambda $, defined in (\ref%
{nonid.MHD.0011.mp2}), removing it from the equations of motion).

The symmetric structure $\left( f,g\right) $ is referred to as \textit{metric%
} component of the motion, and is chosen so that $H$ is conserved during the
motion (\ref{nonid.MHD.0014.mp1}): due to (\ref{nonid.MHD.0009.Casimir}) and
(\ref{nonid.MHD.0011.mp2}), this can be realized by defining $\left(
f,g\right) $ so that%
\begin{equation}
\begin{array}{ccc}
\left( H,f\right) =0 & \forall & f.%
\end{array}
\label{nonid.MHD.0032.mp6}
\end{equation}%
With all these conditions, it's easy to observe the separation of the
metriplectic motion (\ref{nonid.MHD.0014.mp1}) into a Hamiltonian component
plus a metric one: $\left\langle \left\langle f,F\right\rangle \right\rangle
=\left\{ f,H\right\} +\lambda \left( f,C\right) $. The metriplectic
evolution then reads:%
\begin{equation}
\dot{f}=\left\{ f,H\right\} +\lambda \left( f,C\right) .
\label{nonid.MHD.0025.mp3}
\end{equation}%
While the Hamiltonian is conserved due to (\ref{nonid.MHD.0025.mp3}) and (%
\ref{nonid.MHD.0032.mp6}) (completeness of the system $\dot{H}=0$), the
Casimir $C$ chosen in (\ref{nonid.MHD.0011.mp2}) to mimic the entropy
undergoes a non-trivial evolution:%
\begin{equation}
\dot{C}=\lambda \left( C,C\right) .  \label{nonid.MHD.0000.mp4}
\end{equation}%
Due to the semi-definiteness of $\left( f,g\right) $, $\dot{C}$ has a
constant sign: constructing this $C$ as suitably limited from above or
below, it can be used as a Lyapunov quantity for the dynamics (\ref%
{nonid.MHD.0014.mp1}), admitting asymptotic equilibria, as it must be the
case for dissipative systems. The entropic meaning of $C$ will be discussed
more deeply in forthcoming papers. Note, however, that its equation of
motion (\ref{nonid.MHD.0000.mp4}) should be interpreted as an \textit{%
H-Theorem for the MSTDOF} involved in dissipation: in this sense, the
metriplectic scheme represents a simple strategy towards the \textit{%
algebrization of irreversibility}.

\section{Metriplectic formulation of visco-resistive MHD}

The system we want to deal with here is a fully ionized plasma interacting
with the magnetic field generated by its own motion; dissipation takes place
due to the \textit{finite viscosity and resistivity} of the fluid \cite%
{Ray.Choudhuri.book}. More, heat conductivity is finite, hence nearby
parcels of fluid tend to thermalize.

The configuration of the system is given by assigning the bulk velocity $%
\vec{v}$ of the fluid, the magnetic induction $\vec{B}$, the matter density $%
\rho $. Then, another field is introduced expressing the thermodynamical
nature of the matter involved, e.g. the mass-specific entropy density $s$.
The resulting system of equations may be written in an $SO\left( 3\right) $%
-covariant form as:%
\begin{equation}
\left\{ 
\begin{array}{l}
\partial _{t}v_{i}=-v_{k}\partial ^{k}v_{i}-\dfrac{1}{\rho }\partial _{i}p-%
\dfrac{1}{2\rho }\partial _{i}B^{2}+\dfrac{1}{\rho }B_{k}\partial
^{k}B_{i}-\partial _{i}\phi _{grav}+\dfrac{1}{\rho }\partial ^{k}\sigma
_{ik}, \\ 
\\ 
\partial _{t}B_{i}=B_{j}\partial ^{j}v_{i}-B_{i}\partial
^{j}v_{j}-v_{j}\partial ^{j}B_{i}+\mu \partial ^{2}B_{i}, \\ 
\\ 
\partial _{t}\rho =-\partial ^{k}\left( \rho v_{k}\right) , \\ 
\\ 
\partial _{t}s=-v_{k}\partial ^{k}s+\dfrac{\sigma _{ik}}{\rho T}\partial
^{k}v^{i}+\dfrac{\mu }{\rho T}\epsilon _{ikh}\epsilon ^{h}{}_{mn}\partial
^{i}B^{k}\partial ^{m}B^{n}+\dfrac{\kappa }{\rho T}\partial ^{2}T, \\ 
\\ 
\begin{array}{ccc}
\forall & \vec{x}\in \mathbb{D}, & t\in I%
\end{array}%
\end{array}%
\right.  \label{nonid.MHD.0006.nonideal.MHD.system}
\end{equation}%
(here $\mathbb{D\subseteq R}^{3}$ is the space domain where the dynamical
variables are defined and $I\subseteq \mathbb{R}$ is the time interval of
interest). Local thermal equilibrium is assumed, so that the smooth field $T$
may be defined. $\phi _{grav}$ is the gravitational potential to which the
plasma undergoes. The stress tensor $\sigma _{ik}$ is chosen to be linear in
the gradient of the velocity:%
\begin{equation}
\left\{ 
\begin{array}{l}
\sigma _{ik}=\Lambda _{ikmn}\partial ^{m}v^{n}, \\ 
\\ 
\Lambda _{ikmn}\overset{\mathrm{def}}{=}\eta \left( \delta _{ni}\delta
_{mk}+\delta _{nk}\delta _{mi}-\frac{2}{3}\delta _{ik}\delta _{mn}\right)
+\zeta \delta _{ik}\delta _{mn},%
\end{array}%
\right.  \label{nonid.MHD.0030.Lambda}
\end{equation}%
The addendum $\frac{\mu }{\rho T}\epsilon _{ikh}\epsilon ^{h}{}_{mn}\partial
^{i}B^{k}\partial ^{m}B^{n}$ in the fourth equation of (\ref%
{nonid.MHD.0006.nonideal.MHD.system}) is the entropy production rate $\left(
\partial _{t}s\right) _{B}$ due to the Joule Effect and may be obtained
through some considerations of elementary Thermodynamics. In that expression 
$\mu$ indicates the plasma resistivity. The system (\ref%
{nonid.MHD.0006.nonideal.MHD.system}) is \textquotedblleft
closed\textquotedblright\ expressing the quantities $p$ and $T$ in terms of
mass-specific internal energy of the fluid $U$:%
\begin{equation}
\begin{array}{cc}
p=\rho ^{2}\dfrac{\partial U}{\partial \rho }, & T=\dfrac{\partial U}{%
\partial s}.%
\end{array}
\label{nonid.MHD.0007}
\end{equation}%
In the system at hand, the fields $\vec{v}$, $\vec{B}$ and $\rho $ may be
intended as macroscopic, deterministically treated variables, while the
Statistical Mechanics of the MSTDOF giving rise to dissipation is encoded in 
$s$.

The description of the isolated visco-resistive MHD as a complete system 
\cite{Morrison84.001} is possible if the \textquotedblleft total
energy\textquotedblright 
\begin{equation}
H=\dint\limits_{\mathbb{D}}\left[ \dfrac{\rho }{2}v^{2}+\rho \phi _{grav}+%
\dfrac{B^{2}}{2}+\rho U\left( \rho ,s\right) \right] d^{3}x
\label{nonid.MHD.0002.H}
\end{equation}%
is introduced. Thanks to the way in which the Joule Effect contribution
appears in $\partial _{t}s$, it is possible to show that this $H$ is a
constant of motion for the equations (\ref%
{nonid.MHD.0006.nonideal.MHD.system}), provided \textit{suitably good
boundary conditions} are given to the plasma. Indeed, along the motion (\ref%
{nonid.MHD.0006.nonideal.MHD.system}) the quantity $H$ changes only via a
boundary term: $\dot{H}\overset{\partial }{=}0$ ($a\overset{\partial }{=}b$
means that $a$ and $b$ only differ by a boundary term). The
\textquotedblleft suitable conditions\textquotedblright\ at $\partial 
\mathbb{D}$ are those rendering the magnetized plasma an isolated system.

$H$ may be used as the Hamiltonian component of the free energy of the
system which will metriplecticly generate the evolution (\ref%
{nonid.MHD.0006.nonideal.MHD.system}).

Setting to zero the coefficients $\kappa $, $\mu $, $\eta $ and $\zeta $,
the ideal MHD is obtained:%
\begin{equation}
\left\{ 
\begin{array}{l}
\partial _{t}v_{i}=-v_{k}\partial ^{k}v_{i}-\dfrac{1}{\rho }\partial _{i}p-%
\dfrac{1}{2\rho }\partial _{i}B^{2}+\dfrac{1}{\rho }B_{k}\partial
^{k}B_{i}-\partial _{i}\phi _{grav}, \\ 
\\ 
\partial _{t}B_{i}=B_{j}\partial ^{j}v_{i}-B_{i}\partial
^{j}v_{j}-v_{j}\partial ^{j}B_{i}, \\ 
\\ 
\partial _{t}\rho =-\partial ^{k}\left( \rho v_{k}\right) , \\ 
\\ 
\partial _{t}s=-v_{k}\partial ^{k}s.%
\end{array}%
\right.  \label{nonid.MHD.0057.MHD.ideale}
\end{equation}%
The functional $H$ in (\ref{nonid.MHD.0002.H}) is the Hamiltonian for this
field theory \cite{LaJolla.81}, with the noncanonical Poisson bracket \cite%
{Morrison.PRL.noncanonical.ideal.MHD}%
\begin{equation}
\begin{array}{l}
\left\{ f,g\right\} =-\dint\limits_{\mathbb{D}}d^{3}x\left[ \dfrac{\delta f}{%
\delta \rho }\partial _{i}\left( \dfrac{\delta g}{\delta v_{i}}\right) +%
\dfrac{\delta g}{\delta \rho }\partial _{i}\left( \dfrac{\delta f}{\delta
v_{i}}\right) -\dfrac{1}{\rho }\dfrac{\delta f}{\delta v_{i}}\epsilon
_{ikj}\epsilon ^{jmn}\dfrac{\delta g}{\delta v_{k}}\partial _{m}v_{n}+\right.
\\ 
\\ 
+\dfrac{1}{\rho }\dfrac{\delta f}{\delta v_{i}}\epsilon _{ijk}\epsilon
^{kmn}B^{j}\partial _{m}\left( \dfrac{\delta g}{\delta B^{n}}\right) +\dfrac{%
\delta f}{\delta B_{i}}\epsilon _{ijk}\partial ^{j}\left( \dfrac{1}{\rho }%
\epsilon ^{kmn}B_{m}\dfrac{\delta g}{\delta v^{n}}\right) + \\ 
\\ 
\left. +\dfrac{1}{\rho }\partial _{i}s\left( \dfrac{\delta f}{\delta s}%
\dfrac{\delta g}{\delta v_{i}}-\dfrac{\delta g}{\delta s}\dfrac{\delta f}{%
\delta v_{i}}\right) \right] .%
\end{array}
\label{nonid.MHD.0033.symplectic.bracket}
\end{equation}%
Any quantity $f$ is evolved along the motion (\ref{nonid.MHD.0057.MHD.ideale}%
) via the prescription $\dot{f}=\left\{ f,H\right\} $. The Poisson bracket (%
\ref{nonid.MHD.0033.symplectic.bracket}) has several Casimir observables, in
particular we quote those of the form%
\begin{equation}
C\left[ \rho ,s\right] =\dint\limits_{\mathbb{D}}\rho \varphi \left(
s\right) d^{3}x,  \label{nonid.MHD.0001.15.bis}
\end{equation}%
among which one may recognize the total mass $M$ and the total entropy $S$
of the fluid:%
\begin{equation}
\begin{array}{cc}
M\left[ \rho \right] =\dint\limits_{\mathbb{D}}\rho d^{3}x, & S\left[ \rho ,s%
\right] =\dint\limits_{\mathbb{D}}\rho sd^{3}x.%
\end{array}
\label{nonid.MHD.0060.Casimir.M.S}
\end{equation}%
$M$ and $S$ are conserved along the motion (\ref{nonid.MHD.0057.MHD.ideale}%
), because they have zero Poisson bracket with any quantity $f$, and in
particular with $H$. The functionals $C$ in (\ref{nonid.MHD.0001.15.bis})
may be used to construct a metriplectic framework with $H$ in (\ref%
{nonid.MHD.0002.H}), as prescribed in (\ref{nonid.MHD.0011.mp2}) and (\ref%
{nonid.MHD.0014.mp1}).

Other non-Casimir quantities, remarkably conserved by the motion (\ref%
{nonid.MHD.0057.MHD.ideale}), are all the space-time symmetries related to
the Galileo transformation, i.e. the total momentum $\vec{P}$ of the system,
the total angular momentum $\vec{L}$ and a quantity $\vec{G}$ which is the
symplectic generator of Galileo's boosts. Their definitions%
\begin{equation}
\left\{ 
\begin{array}{l}
\begin{array}{cc}
P_{h}=\dint\limits_{\mathbb{D}}\rho v_{h}d^{3}x, & L_{h}=\dint\limits_{%
\mathbb{D}}\rho \epsilon _{hij}x^{i}v^{j}d^{3}x,%
\end{array}
\\ 
\\ 
G_{h}=\dint\limits_{\mathbb{D}}\rho \left( x_{h}-v_{h}t\right) d^{3}x%
\end{array}%
\right.  \label{nonid.MHD.0062.Galileo.Group}
\end{equation}%
plus the definition of $H$ in (\ref{nonid.MHD.0002.H}) and of the Poisson
bracket $\left\{ f,g\right\} $ in (\ref{nonid.MHD.0033.symplectic.bracket})
imply:%
\begin{equation}
\begin{array}{ccc}
\left\{ P_{h},H\right\} \overset{\partial }{=}0, & \left\{ L_{h},H\right\} 
\overset{\partial }{=}0, & \left\{ G_{h},H\right\} \overset{\partial }{=}0.%
\end{array}
\label{nonid.MHD.0063}
\end{equation}

Let's turn back to the system with dissipation (\ref%
{nonid.MHD.0006.nonideal.MHD.system}): the dissipative terms appearing there
must be given by a suitable symmetric bracket $\left( f,g\right) $ (still to
be defined) of the dynamical variables at hand with the Casimir $C$ to be
used as in (\ref{nonid.MHD.0025.mp3}). The correct Casimir to be used is the
plasma entropy $S\left[ \rho ,s\right] $ in (\ref{nonid.MHD.0060.Casimir.M.S}%
). \textit{The result presented here is the explicit expression of such
bracket} $\left( f,g\right) $.

The dissipative terms in (\ref{nonid.MHD.0006.nonideal.MHD.system}) are the
8 expressions%
\begin{equation*}
\begin{array}{cccc}
D_{i}^{\left( v\right) }=\dfrac{1}{\rho }\partial ^{k}\sigma _{ik}, & 
D_{i}^{\left( B\right) }=\mu \partial ^{2}B_{i}, & D^{\left( \rho \right)
}=0, & D^{\left( s\right) }=\dfrac{1}{\rho T}\left( \sigma _{ik}\partial
^{k}v^{i}+\mu j^{2}+\kappa \partial ^{2}T\right) ,%
\end{array}%
\end{equation*}%
with self-evident meaning of the symbols. If these terms are collected in an
8-uple $\underline{D}=\left( \vec{D}^{\left( v\right) },\vec{D}^{\left(
B\right) },D^{\left( \rho \right) },D^{\left( s\right) }\right) $ and the
dynamical variables in (\ref{nonid.MHD.0006.nonideal.MHD.system}) are $%
\underline{\psi }=\left( \vec{v},\vec{B},\rho ,s\right) $, then one aims to
define the metric bracket $\left( f,g\right) $ so that $\underline{D}%
=\lambda \left( \underline{\psi },S\right) $.

Since the metriplectic scheme for a dissipative neutral fluid has been
already worked out in \cite{Morrison84.001}, here $\left( f,g\right) $ for
the system (\ref{nonid.MHD.0006.nonideal.MHD.system}) will be defined by
generalizing the expressions of the metric part of dynamics to include the
Joule effect dissipation. Considering (\ref{nonid.MHD.0030.Lambda}), the
dissipation element in the $\vec{v}$-equation and the corresponding entropy
production due to the velocity gradients \ show a beautiful parallel with
the same terms pertaining to the motion of $\vec{B}$:%
\begin{equation*}
\left\{ 
\begin{array}{l}
\begin{array}{ccc}
D_{i}^{\left( v\right) }=\dfrac{1}{\rho }\partial ^{k}\left( \Lambda
_{kimn}\partial ^{m}v^{n}\right) , & \; & D_{i}^{\left( B\right) }=\partial
^{k}\left( \Theta _{kimn}\partial ^{m}B^{n}\right) ,%
\end{array}
\\ 
\\ 
\Theta _{jkmn}\overset{\mathrm{def}}{=}\mu \epsilon _{jki}\epsilon
^{i}{}_{mn}, \\ 
\\ 
\begin{array}{ccc}
\left( \partial _{t}s\right) _{v}=\dfrac{1}{\rho T}\Lambda _{jkmn}\partial
^{j}v^{k}\partial ^{m}v^{n}, & \; & \left( \partial _{t}s\right) _{B}=\dfrac{%
1}{\rho T}\Theta _{jkmn}\partial ^{j}B^{k}\partial ^{m}B^{n}.%
\end{array}%
\end{array}%
\right.
\end{equation*}%
The system (\ref{nonid.MHD.0006.nonideal.MHD.system}) may be re-written as:%
\begin{equation}
\left\{ 
\begin{array}{l}
\partial _{t}v_{i}=-v_{k}\partial ^{k}v_{i}-\dfrac{1}{\rho }\partial _{i}p-%
\dfrac{1}{2\rho }\partial _{i}B^{2}+\dfrac{1}{\rho }B_{k}\partial
^{k}B_{i}-\partial _{i}\phi _{grav}+\dfrac{1}{\rho }\partial ^{k}\left(
\Lambda _{kimn}\partial ^{m}v^{n}\right) , \\ 
\\ 
\partial _{t}B_{i}=B_{j}\partial ^{j}v_{i}-B_{i}\partial
^{j}v_{j}-v_{j}\partial ^{j}B_{i}+\partial ^{k}\left( \Theta _{kimn}\partial
^{m}B^{n}\right) , \\ 
\\ 
\partial _{t}\rho =-\partial ^{k}\left( \rho v_{k}\right) , \\ 
\\ 
\partial _{t}s=-v_{k}\partial ^{k}s+\dfrac{1}{\rho T}\Lambda _{kimn}\partial
^{k}v^{i}\partial ^{m}v^{n}+\dfrac{1}{\rho T}\Theta _{kimn}\partial
^{k}B^{i}\partial ^{m}B^{n}+\dfrac{\kappa }{\rho T}\partial ^{2}T.%
\end{array}%
\right.  \label{nonid.MHD.0018.pde.symm.v.B}
\end{equation}%
In both the cases of $\vec{v}$ and of $\vec{B}$, the dissipative term is
given by the divergence of the contraction of a rank-4-tensor ($\Lambda
_{kimn}$ and $\Theta _{kimn}$ respectively) with the gradient of the local
variable ($\partial ^{m}v^{n}$ and $\partial ^{m}B^{n}$ respectively); in
both the cases, the contribution to the entropy production is \textit{a
quadratic form in the gradients of the field}, $\frac{1}{\rho T}\Lambda
_{kimn}\partial ^{k}v^{i}\partial ^{m}v^{n}$ and $\frac{1}{\rho T}\Theta
_{kimn}\partial ^{k}B^{i}\partial ^{m}B^{n}$ respectively (\textit{quadratic
dissipation}). In \cite{Morrison84.001} the dissipative part of the motion
of a viscous Navier-Stokes system is accounted for via%
\begin{equation}
\begin{array}{l}
\left( f,g\right) _{\mathrm{NS}}=\dfrac{1}{\lambda }\dint\limits_{\mathbb{D}%
}d^{3}x\left\{ T\Lambda _{ikmn}\left[ \partial ^{i}\left( \dfrac{1}{\rho }%
\dfrac{\delta f}{\delta v_{k}}\right) -\dfrac{1}{\rho T}\partial ^{i}v^{k}%
\dfrac{\delta f}{\delta s}\right] \left[ \partial ^{m}\left( \dfrac{1}{\rho }%
\dfrac{\delta g}{\delta v_{n}}\right) -\dfrac{1}{\rho T}\partial ^{m}v^{n}%
\dfrac{\delta g}{\delta s}\right] +\right. \\ 
\\ 
\left. +\kappa T^{2}\partial ^{k}\left( \dfrac{1}{\rho T}\dfrac{\delta f}{%
\delta s}\right) \partial _{k}\left( \dfrac{1}{\rho T}\dfrac{\delta g}{%
\delta s}\right) \right\} :%
\end{array}
\label{nonid.MHD.0019.NS.metric}
\end{equation}%
the addendum linear in $\Lambda _{ikmn}$ accounts for the dissipation as in
the equations of motion of $\vec{v}$ and for the entropy production due to
the viscosity. The other addendum describes the entropy variation due to the
heat transport. The analogy between the quadratic dissipation for $\vec{v}$
and that for $\vec{B}$ suggests that the bracket for the dissipative MHD
should be of the form:%
\begin{equation}
\begin{array}{l}
\left( f,g\right) =\dfrac{1}{\lambda }\dint\limits_{\mathbb{D}}d^{3}x\left\{
T\Lambda _{ikmn}\left[ \partial ^{i}\left( \dfrac{1}{\rho }\dfrac{\delta f}{%
\delta v_{k}}\right) -\dfrac{1}{\rho T}\partial ^{i}v^{k}\dfrac{\delta f}{%
\delta s}\right] \left[ \partial ^{m}\left( \dfrac{1}{\rho }\dfrac{\delta g}{%
\delta v_{n}}\right) -\dfrac{1}{\rho T}\partial ^{m}v^{n}\dfrac{\delta g}{%
\delta s}\right] +\right. \\ 
\\ 
+T\Theta _{ikmn}\left[ \partial ^{i}\left( \dfrac{\delta f}{\delta B_{k}}%
\right) -\dfrac{1}{\rho T}\partial ^{i}B^{k}\dfrac{\delta f}{\delta s}\right]
\left[ \partial ^{m}\left( \dfrac{\delta g}{\delta B_{n}}\right) -\dfrac{1}{%
\rho T}\partial ^{m}B^{n}\dfrac{\delta g}{\delta s}\right] + \\ 
\\ 
\left. +\kappa T^{2}\partial ^{k}\left( \dfrac{1}{\rho T}\dfrac{\delta f}{%
\delta s}\right) \partial _{k}\left( \dfrac{1}{\rho T}\dfrac{\delta g}{%
\delta s}\right) \right\} .%
\end{array}
\label{nonid.MHD.0026.try.bracket.1}
\end{equation}%
This bracket is shown to be the right one to produce the dissipative terms
in (\ref{nonid.MHD.0018.pde.symm.v.B}) once the free energy is chosen as $%
F=H+\lambda S$, $H$ being the Hamiltonian defined in (\ref{nonid.MHD.0002.H}%
) and $S$ the total entropy given in (\ref{nonid.MHD.0060.Casimir.M.S}), so
that:%
\begin{equation}
F\left[ \vec{v},\vec{B},\rho ,s\right] =\dint\limits_{\mathbb{D}}\left[ 
\dfrac{\rho }{2}v^{2}+\rho \phi _{grav}+\dfrac{B^{2}}{2}+\rho U\left( \rho
,s\right) +\lambda \rho s\right] d^{3}x.  \label{nonid.MHD.0066.free.energy}
\end{equation}

The metric bracket (\ref{nonid.MHD.0026.try.bracket.1}) is shown to generate
the dissipative part of $\partial _{t}\vec{v}$, because this $\left(
f,g\right) $ is exactly the $\left( f,g\right) _{\mathrm{NS}}$ in (\ref%
{nonid.MHD.0019.NS.metric}) for the part concerning the velocity field; the
addendum involving $\frac{\delta }{\delta \vec{B}}$ does not contribute to $%
\partial _{t}\vec{v}$. It contributes instead to the dissipative part of $%
\partial _{t}\vec{B}$, calculated as $\lambda \left( B_{h},S\right) =\mu
\partial ^{2}B_{h}$.

The bracket in (\ref{nonid.MHD.0026.try.bracket.1}) is symmetric in the
exchange $f\leftrightarrow g$, due to the property $\Lambda _{ikmn}=\Lambda
_{mnik}$ and $\Theta _{ikmn}=\Theta _{mnik}$, and the self-evident symmetry
of the addendum $\kappa T^{2}\partial ^{k}\left( \frac{1}{\rho T}\frac{%
\delta f}{\delta s}\right) \partial _{k}\left( \frac{1}{\rho T}\frac{\delta g%
}{\delta s}\right) $. As far as its semi-definiteness is concerned, consider
that it has been constructed by summing the bracket $\left( f,g\right) _{%
\mathrm{NS}}$ in (\ref{nonid.MHD.0019.NS.metric}) and the bracket%
\begin{equation}
\left( f,g\right) _{B}=\dfrac{1}{\lambda }\dint\limits_{\mathbb{D}%
}d^{3}xT\Theta _{ikmn}\left[ \partial ^{i}\left( \dfrac{\delta f}{\delta
B_{k}}\right) -\dfrac{1}{\rho T}\partial ^{i}B^{k}\dfrac{\delta f}{\delta s}%
\right] \left[ \partial ^{m}\left( \dfrac{\delta g}{\delta B_{n}}\right) -%
\dfrac{1}{\rho T}\partial ^{m}B^{n}\dfrac{\delta g}{\delta s}\right] .
\label{nonid.MHD.0041.bracket.J}
\end{equation}%
The semi-definiteness of $\left( f,g\right) _{\mathrm{NS}}$ was proved in 
\cite{Morrison84.001}. The newly added Joule term $\left( f,g\right) _{B}$,
for which one has:%
\begin{equation*}
\left\{ 
\begin{array}{l}
\left( f,f\right) _{B}=\dfrac{1}{\lambda }\dint\limits_{\mathbb{D}%
}d^{3}xT\Theta _{ikmn}T^{ik}\left( f\right) T^{mn}\left( f\right) , \\ 
\\ 
T^{ab}\left( f\right) =\partial ^{a}\left( \dfrac{\delta f}{\delta B_{b}}%
\right) -\dfrac{1}{\rho T}\partial ^{a}B^{b}\dfrac{\delta f}{\delta s}.%
\end{array}%
\right.
\end{equation*}%
$T^{ab}\left( f\right) $ can be subdivided into a symmetric part $%
S^{ab}\left( f\right) =\frac{1}{2}\left[ T^{ab}\left( f\right) +T^{ba}\left(
f\right) \right] $ plus an antisymmetric part $A^{ab}\left( f\right) =\frac{1%
}{2}\left[ T^{ab}\left( f\right) -T^{ba}\left( f\right) \right] $, and, due
to the symmtry properties of $\Theta _{ikmn}$, $\Theta _{ikmn}=-\Theta
_{kimn}$ and $\Theta _{ikmn}=-\Theta _{iknm}$, one can replace $T^{ab}\left(
f\right) $ with its antisymmetric part $A^{ab}\left( f\right) $ only, since
the symmetric parts will be canceled in the calculation of $\left(
f,f\right) _{B}$:%
\begin{equation*}
\left( f,f\right) _{B}=\dfrac{2}{\lambda }\dsum\limits_{i,k}\dint\limits_{%
\mathbb{D}}\mu TA_{ik}^{2}\left( f\right) d^{3}x.
\end{equation*}%
The sign of this expression is just that of $\lambda $ for every functional $%
f$. The semi-definiteness of the whole $\left( f,g\right) =\left( f,g\right)
_{\mathrm{NS}}+\left( f,g\right) _{B}$ is proved (so that $S$ may be
considered a good Lyapunov functional).

Last but not least, the metric algebra (\ref{nonid.MHD.0026.try.bracket.1})
generates exactly the local entropy production due to the mechanisms of
dissipation and heat transport: $\lambda \left( s,S\right) =D^{\left(
s\right) }$.

It is possible to show that the functional gradient of the Hamiltonian is a
null mode of the metric algebra (\ref{nonid.MHD.0026.try.bracket.1}):%
\begin{equation*}
\begin{array}{ccc}
\left( H,f\right) =0 & \forall & f.%
\end{array}%
\end{equation*}%
Also, the metric part of the motion algebra keeps the quantities in (\ref%
{nonid.MHD.0062.Galileo.Group}) constant:%
\begin{equation}
\begin{array}{ccc}
\left( P_{h},S\right) =0, & \left( L_{h},S\right) =0, & \left(
G_{h},S\right) =0.%
\end{array}
\label{nonid.MHD.0010.25.bis}
\end{equation}%
Equation (\ref{nonid.MHD.0010.25.bis}), together with (\ref{nonid.MHD.0063}%
), renders the metriplectic motion of the non-ideal MHD invariant under the
transformations of the Galileo Group.

The metriplectic bracket%
\begin{equation}
\begin{array}{l}
\left\langle \left\langle f,g\right\rangle \right\rangle =-\dint\limits_{%
\mathbb{D}}d^{3}x\left[ \dfrac{\delta f}{\delta \rho }\partial _{i}\left( 
\dfrac{\delta g}{\delta v_{i}}\right) +\dfrac{\delta g}{\delta \rho }%
\partial _{i}\left( \dfrac{\delta f}{\delta v_{i}}\right) -\dfrac{1}{\rho }%
\dfrac{\delta f}{\delta v_{i}}\epsilon _{ikj}\epsilon ^{jmn}\dfrac{\delta g}{%
\delta v_{k}}\partial _{m}v_{n}+\right. \\ 
\\ 
+\dfrac{1}{\rho }\dfrac{\delta f}{\delta v_{i}}\epsilon _{ijk}\epsilon
^{kmn}B^{j}\partial _{m}\left( \dfrac{\delta g}{\delta B^{n}}\right) +\dfrac{%
\delta f}{\delta B_{i}}\epsilon _{ijk}\partial ^{j}\left( \dfrac{1}{\rho }%
\epsilon ^{kmn}B_{m}\dfrac{\delta g}{\delta v^{n}}\right) + \\ 
\\ 
\left. +\dfrac{1}{\rho }\partial _{i}s\left( \dfrac{\delta f}{\delta s}%
\dfrac{\delta g}{\delta v_{i}}-\dfrac{\delta g}{\delta s}\dfrac{\delta f}{%
\delta v_{i}}\right) \right] +\dfrac{1}{\lambda }\dint\limits_{\mathbb{D}%
}d^{3}xT\left\{ \kappa T\partial ^{k}\left( \dfrac{1}{\rho T}\dfrac{\delta f%
}{\delta s}\right) \partial _{k}\left( \dfrac{1}{\rho T}\dfrac{\delta g}{%
\delta s}\right) +\right. \\ 
\\ 
+\Lambda _{ikmn}\left[ \partial ^{i}\left( \dfrac{1}{\rho }\dfrac{\delta f}{%
\delta v_{k}}\right) -\dfrac{1}{\rho T}\partial ^{i}v^{k}\dfrac{\delta f}{%
\delta s}\right] \left[ \partial ^{m}\left( \dfrac{1}{\rho }\dfrac{\delta g}{%
\delta v_{n}}\right) -\dfrac{1}{\rho T}\partial ^{m}v^{n}\dfrac{\delta g}{%
\delta s}\right] + \\ 
\\ 
\left. +\Theta _{ikmn}\left[ \partial ^{i}\left( \dfrac{\delta f}{\delta
B_{k}}\right) -\dfrac{1}{\rho T}\partial ^{i}B^{k}\dfrac{\delta f}{\delta s}%
\right] \left[ \partial ^{m}\left( \dfrac{\delta g}{\delta B_{n}}\right) -%
\dfrac{1}{\rho T}\partial ^{m}B^{n}\dfrac{\delta g}{\delta s}\right]
\right\} ,%
\end{array}
\label{nonid.MHD.0004.metriplectic.bracket}
\end{equation}%
obtained by putting together (\ref{nonid.MHD.0033.symplectic.bracket}) and (%
\ref{nonid.MHD.0026.try.bracket.1}), has all the features required to govern
the visco-resistive MHD, with the free energy defined in (\ref%
{nonid.MHD.0066.free.energy}).

As suggested in \cite{morrison.paradigm}, it is possible to determine the
equilibrium configurations by studying the extrema of the free energy $F$.
The functional derivatives of $F$ read%
\begin{equation*}
\left\{ 
\begin{array}{l}
\begin{array}{cc}
\dfrac{\delta F}{\delta \vec{v}}=\rho \vec{v}, & \dfrac{\delta F}{\delta 
\vec{B}}=\vec{B},%
\end{array}
\\ 
\\ 
\dfrac{\delta F}{\delta \rho }=\dfrac{v^{2}}{2}+\phi _{grav}+U+\rho \dfrac{%
\partial U}{\partial \rho }+\lambda s, \\ 
\\ 
\dfrac{\delta F}{\delta s}=\rho \dfrac{\partial U}{\partial s}+\lambda \rho ,%
\end{array}%
\right.
\end{equation*}%
so that, setting them to zero and considering the thermodynamical closure (%
\ref{nonid.MHD.0007}), the asymptotic equilibrium configuration is found to
be:%
\begin{equation}
\left\{ 
\begin{array}{l}
\begin{array}{ccc}
\vec{v}_{\mathrm{eq}}=0, & \vec{B}_{\mathrm{eq}}=0, & T_{\mathrm{eq}%
}=-\lambda ,%
\end{array}
\\ 
\\ 
p_{\mathrm{eq}}+\rho _{\mathrm{eq}}\phi _{grav}=\rho _{\mathrm{eq}}\left(
Ts-U\right) _{\mathrm{eq}}.%
\end{array}%
\right.  \label{nonid.MHD.0008.eq}
\end{equation}%
A configuration towards which the system may tend\ to relax (under suitable
initial conditions) has neither bulk velocity, nor magnetic induction, while
pressure and external forces equilibrate the thermodynamical free energy of
the gas, and the temperature field matches everywhere minus the constant $%
\lambda $. At the equilibrium, the free energy of the metriplectic scheme
really appears to be isomorphic to the expression known in traditional
Thermodynamics $F=H-T_{\mathrm{eq}}S$, being $H$ the energy of the fluid and 
$S$ its entropy.

As a corollary of the above results, one can obtain the metriplectic
formulation of reduced MHD models \cite{biskamp.nonlinear.book}, which are
widely used when the dependence on one of the spatial coordinates can be
ignored. This can be the case, form instance, of tokamak fusion devices, in
which the presence of a strong toroidal component of the magnetic field $%
\vec{B}_{0}$, makes the dynamics essentially two-dimensional and taking
place on the poloidal plane, perpendicular to the toroidal direction.
Several such examples may be done both in astrophysical plasmas and fusion
plasmas.

An incompressible 2D resistive MHD model, accounting for entropy production,
may be obtained reducing the 3D system, taking the limit of zero viscosity
and adopting magnetic potential, vorticity and entropy per unit mass, as
dynamical variables \cite{referenza.RMHD}:%
\begin{equation}
\left\{ 
\begin{array}{l}
\dfrac{\partial \psi }{\partial t}+[\phi ,\psi ]=\mu \partial _{\bot
}^{2}\psi , \\ 
\\ 
\dfrac{\partial \omega }{\partial t}+[\phi ,\omega ]+[\partial _{\bot
}^{2}\psi ,\psi ]=0, \\ 
\\ 
\dfrac{\partial s}{\partial t}+[\phi ,s]=\dfrac{\mu }{\rho _{0}T}(\partial
_{\bot }^{2}\psi )^{2}.%
\end{array}%
\right.  \label{e1.17}
\end{equation}%
In the above equations $\psi $ is the poloidal magnetic flux, $\phi $ the
stream function, $\omega =\partial _{\bot }^{2}\phi $ the plasma vorticity, $%
s$ the entropy per unit mass, $\mu $ the resistivity and $\left[ a,b\right]
=\partial _{x}a\partial _{y}b-\partial _{y}a\partial _{x}b$ is the canonical
bracket in the $x$, $y$ coordinates along the poloidal plane. $\vec{\partial}%
_{\bot }$ is the gradient along the plane orthogonal to $\vec{B}_{0}$, i.e.
the poloidal plane, and $\partial _{\bot }^{2}$ is the corresponding
Laplacian. All fields depend on $x$ and $y$ only. Consistently with the
incompressibility assumption, the mass density $\rho _{0}$ is taken to be
constant.

Although deprived of the terms depending on the fluid viscosity, the model (%
\ref{e1.17}) is a useful tool for investigating, for instance, the
phenomenon of magnetic reconnection \cite{Bis00,Pri00}, in which the
dissipative term depending on the resistivity, allows for the change of
topology of magnetic field line configurations, in addition to converting
magnetic energy into heat.

The Hamiltonian component of the motions in (\ref{e1.17}), obtained in the
limit $\mu =0$, is generated by the Hamiltonian functional%
\begin{equation}
H=\frac{1}{2}\int d^{2}x(|\vec{\partial}_{\bot }\psi |^{2}+|\vec{\partial}%
_{\bot }\phi |^{2})+\rho _{0}\int d^{2}xU(s)  \label{ham}
\end{equation}%
and by the Poisson bracket%
\begin{equation}
\{f,g\}=\dint d^{2}x\left( \psi ([f_{\psi },g_{\omega }]+[f_{\omega
},g_{\psi }])+\omega \lbrack f_{\omega },g_{\omega }]+s([f_{s},g_{\omega
}]+[f_{\omega },g_{s}])\right) ,  \label{pb}
\end{equation}%
where subscripts indicate functional derivatives.

The last term on the right-hand side of (\ref{ham}) comes from the
contribution of the internal energy $U$. In the constant density limit,
however, such term is actually a Casimir of the bracket (\ref{pb}). The
dissipative part of the system is generated with the help of a metric
bracket $(,)$. In the incomplete case, with no entropy evolution, the
symmetric bracket producing the resistive term in the Ohm's law in (\ref%
{e1.17}), had been presented in \cite{Mor84}. For the above complete system,
the dissipative part is obtained from the $(,)_{B}$ metric bracket presented
in (\ref{nonid.MHD.0041.bracket.J}), by applying the relation $\vec{\partial}%
_{\bot }\times f_{\vec{B}}=f_{\vec{A}}$, where $\vec{A}$ is the magnetic
vector potential and $\vec{B}$ the magnetic induction, and then by
projecting in 2D. The result is%
\begin{equation}
(f,g)=\frac{\mu }{\lambda }\int d^{2}x\left( Tf_{\psi }g_{\psi }+\frac{%
\partial _{\bot }^{2}\psi }{\rho _{0}}(f_{\psi }g_{s}+f_{s}g_{\psi })+\frac{%
(\partial _{\bot }^{2}\psi )^{2}}{\rho _{0}^{2}T}f_{s}g_{s}\right) .
\label{sb}
\end{equation}

For this reduced model, the properties characterizing the metric bracket can
be shown with more immediacy. The bracket (\ref{sb}), indeed, is evidently
symmetric. The relation 
\begin{equation*}
(H,g)=\dfrac{\mu }{\lambda }\dint d^{2}x\left( T(-\partial _{\bot }^{2}\psi
)g_{\psi }+\dfrac{\partial _{\bot }^{2}\psi }{\rho _{0}}(-\partial _{\bot
}^{2}\psi )g_{s}+\rho _{0}Tg_{\psi })+\dfrac{(\partial _{\bot }^{2}\psi )^{2}%
}{\rho _{0}^{2}T}\rho _{0}Tg_{s}\right) =0
\end{equation*}%
shows that the functional gradient of $H$ is in the kernel of the metric
bracket for any $g$. Concerning semi-definiteness one can see that 
\begin{equation*}
(f,f)=\frac{\mu }{\lambda }\int d^{2}xT\left( f_{\psi }+\frac{\partial
_{\bot }^{2}\psi }{\rho _{0}T}f_{s}\right) ^{2},
\end{equation*}%
so that $(f,f)$ has the same sign of $\lambda $. Finally, upon defining%
\begin{equation*}
F=H+\lambda \rho _{0}\int sd^{2}x,
\end{equation*}%
one can verify that $(\psi ,F)$, $(\omega ,F)$ and $(s,F)$ yield the desired
dissipative terms:%
\begin{equation*}
\begin{array}{l}
\begin{array}{cc}
(\psi ,F)=\mu \partial _{\bot }^{2}\psi , \qquad (\omega ,F)=0, & 
\end{array}
\\ 
\\ 
(s,F) =\dfrac{\mu }{\rho _{0}T}(\partial _{\bot }^{2}\psi )^{2}.%
\end{array}%
\end{equation*}

\section{Conclusions}

The metriplectic formulation of the visco-resistive MHD equations has been
derived. Such formulation is identified by a free energy functional, given
by the sum of the Hamiltonian of ideal MHD with the entropy Casimir, and a
bracket obtained by summing the Poisson bracket of ideal MHD with a new
metric bracket yielding the dissipative terms. The metric bracket extends
that derived in Ref. \cite{Morrison84.001} for the Navier-Stokes equations.
In addition to yielding the desired dissipative terms, the bracket is shwon
to conserve the Hamiltonian of ideal MHD as well as other constants of
motion of the ideal limit, related to space-time symmetries. The dynamics
governed by this metriplectic system is then shown to tend asymptotically in
time, toward states with no flow and no magnetic energy. From the general
results on visco-resistive MHD, we obtained also the metriplectic
formulation of a reduced resistive model for incompressible plasmas.

Concerning future directions, some equilibrium configuration less trivial
than (\ref{nonid.MHD.0008.eq}) should be investigated: the configuration (%
\ref{nonid.MHD.0008.eq}) is \textquotedblleft entropic
death\textquotedblright , taking place when friction has dissipated all the
bulk kinetic and magnetic energy into heat. The existence of the equilibrium
configuration (\ref{nonid.MHD.0008.eq}) is very intuitive, it is a
configuration reachable from initial zero Galileo charges (\ref%
{nonid.MHD.0062.Galileo.Group}), but it represents only one possible final
state. Actually, even if the free energy (\ref{nonid.MHD.0066.free.energy})
seems to predict only this equilibrium configuration, other relaxation
plasma states are known in nature, justifiable in this framework by
generalizing the functional $F$ in (\ref{nonid.MHD.0066.free.energy}) to
some $F^{\prime }$ in which constraints not considered here are brought into
the play. An extremization of $F$ conditioned to initial values of the
quantities in (\ref{nonid.MHD.0062.Galileo.Group}) would, for instance, give
a final $\vec{v}$ different from zero. Even more interesting would be the
extension of $F$ to expressions in which the Casimir functional $C$ in (\ref%
{nonid.MHD.0011.mp2}) is not simply restricted to $S$, but involves the
velocity and the magnetic degrees of freedom \cite{Casimir.MHD.ideale}.

All the conditioning schemes just mentioned appear to be very smart, but
should better be deduced from a consistent \textquotedblleft First
Principle\textquotedblright\ of metriplectic Physics, which is not yet clear
to the Authors.

A second important remark, is that the relationship between $S$ in the
evolution of the dissipative system, and its information theory
interpretation should be investigated. Indeed, on the one hand, the
relationship (\ref{nonid.MHD.0025.mp3}) renders $S$ a piece of the
functional $F$ that \textit{metriplectically generates the time translations}%
, so that the entropy is recognized as the quantity that is \textit{fully
responsible for dissipation}. On the other hand, $S$ should quantify the
lack of information about the precise state of the MSTDOF: in the
metriplectic scheme, however, no mention to probability is done, it is
apparently a fully deterministic dynamics, even if the proper Thermodynamics
emerges clearly. The metriplectic framework could probably emerge in a
natural way within the Physics of a \textit{Hamiltonian system} interacting
with \textit{noise}, that represents the MSTDOF free to fluctuate
stochastically \cite{phythian}. Such a stochastic scenario is expected to be
approximated by the deterministic dynamics (\ref{nonid.MHD.0025.mp3}) under
suitable hypotheses. The theory of stochastic systems will be of great help
in this line of research \cite{FokkerPlanck.book}.

\bigskip

\bigskip

\textbf{ACKNOWLEDGEMENTS}

\bigskip

The authors are grateful to Philip J. Morrison of the Institute for Fusion
Studies of the University of Texas in Austin, for useful discussions an
criticism; to Cristel Chandre of the Center for Theoretical Physics in
Marseille for useful discussions and for reading the manuscript.

The work of Massimo Materassi has been partially supported by EURATOM, who
contributed through the \textquotedblleft Contratto di Associazione
Euratom-Enea-CNR\textquotedblright , in the framework of the ITER Project.
This work was supported by the European Community under the contracts of
Association between EURATOM, CEA, and the French Research Federation for
fusion studies. The views and opinions expressed herein do not necessarily
reflect those of the European Commission. Financial support was also
received from the Agence Nationale de la Recherche (ANR GYPSI n. 2010 BLAN
941 03) 

\end{document}